\documentclass[review,number,sort&compress]{elsarticle}

\usepackage{lineno,hyperref}
\usepackage{amsmath}
\usepackage{subcaption}
\usepackage{pdflscape} 
\usepackage{units}
\usepackage{gensymb}

\journal{Nucl. Instr. Meth. Phys. Res. -A}

\begin{document}

\begin{frontmatter}

\title{Using a portable muon detector for radioactive source measurements and identification}

\author[pucp]{J. ~Masias}
\author[pucp]{F. ~Delgado}
\author[pucp]{L. ~Coll}
\author[pucp]{A. ~P\'erez}
\author[pucp]{J. ~Bazo \corref{cor1}}
\ead{jbazo@pucp.edu.pe}
\author[pucp]{A.M. ~Gago}

\cortext[cor1]{Corresponding author}
\address[pucp]{Secci\'on F\'isica, Departamento de Ciencias, Pontificia Universidad Cat\'olica del Per\'u, Av. Universitaria 1801, Lima 32, Per\'u}

\begin{abstract}
We have re-purposed portable plastic scintillator muon detectors, designed by the CosmicWatch project, for the measurement of electrons emitted by the decay of radioactive sources. 
For the latter purpose we have first calibrated the detectors using the local atmospheric muon flux, performing angular distribution and attenuation measurements. In addition, we have simulated the detector using Geant4 in a detailed fashion for a cross-check and better understanding of the device.
Then, we have developed a method to evaluate the activity of $\beta$-sources and to discriminate different $\beta$-sources by looking into their respective voltage spectrum output. 
\end{abstract}


\end{frontmatter}


\section{\label{sec:level1}Introduction}

In order to detect charged particles, especially muons, several portable devices have been designed. These devices are based on different technologies such as plastic scintillators \citep{845784, PNNL-20194}, silicon pixels \citep{alice} and drift tubes \citep{NewMexicoTech}. For this work, we have assembled and used hand-held detectors based on plastic scintillator developed by the CosmicWatch project \citep{Axani:2016xur, Axani:2018qzs}.

This detector can measure muons that are produced in the upper layers of the atmosphere. Muon production occurs when cosmic rays, mainly protons, interact with nuclei in the atmosphere, producing secondary particles, which then decay into muons \cite{Gaisser}. Muons are highly penetrating, a property that allows us to detect them under different conditions (e.g. inside buildings or below ground level).

In this paper we will show that this detector, after minor modifications, can be used to count electrons as well. By doing this, we will measure radioactive $\beta$-sources activity and we will study the feasibility of identifying different elements using their measured spectrum. Other plastic scintillator detectors have also been used for $\beta$-sources measurements \citep{KronaThesis, betaray, PlasticBeta}.

This paper is divided as follows: we first describe in Sec. \ref{sec:detector} the detector. Then, in Sec. \ref{sec:simulation} we outline how we performed simulations of the detector under different configurations to cross-check the physics involved. In Sec. \ref{sec:calibration} we describe the calibration process using atmospheric muon measurements and simulations. Then in Sec. \ref{sec:radioactive} we describe the method and give the results on the measurement of the radioactive $\beta$-sources activities and source identification.

\section{Desktop Muon Detector}
\label{sec:detector}
\begin{figure}[htb]
\centering
\includegraphics[scale=0.4]{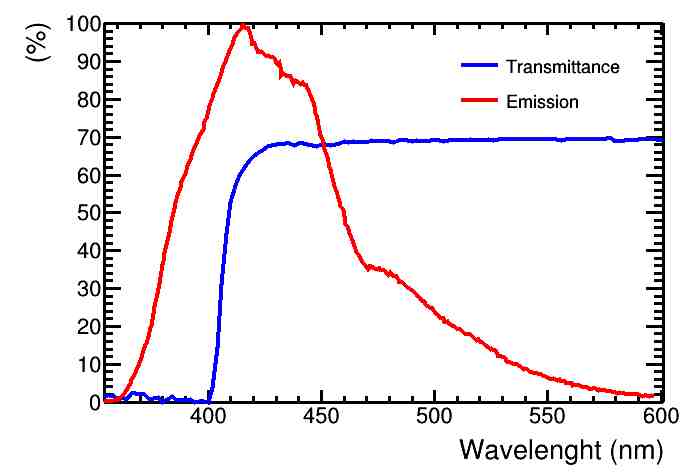}
\caption{Measured transmittance (spectrophotometer) and emission (photoluminescence spectroscopy) for the polystyrene doped with 1\% PPO + 0.03\% POPOP  scintillator. The emission curve has been normalized with its maximum intensity.}
\label{fig:Scintillator_properties}
\end{figure}

The Desktop Muon Detector (DMD), designed by the CosmicWatch project from the Massachusetts Institute of Technology, is a scintillation based device \citep{Axani:2016xur}.	The core of the detector is a plastic scintillator approximately 5 cm $\times$ 5 cm $\times$ 1 cm polystyrene Dow Styron 663 W doped with 1\% PPO + 0.03\% POPOP \cite{PlaDalmau}. We have measured its optical properties, as shown in Fig. \ref{fig:Scintillator_properties}, corroborating that it is blue-emitting (peak emission at $\approx$420 nm) and has an absorption cut-off at $\approx$400 nm. 

When a charged particle passes through the plastic, it leaves a fraction of its energy and the medium emits scintillation light. 
A SensL C-Series SMT silicon photomultiplier (SiPM) \citep{sensL} ($6.0\times6.0$ mm$^2$), attached to the plastic, detects these scintillation photons, with maximum efficiency at 420 nm. To improve the signal response, we use a 29.4 V bias voltage (24.7 V breakdown voltage and 4.7 V overvoltage), which is delivered by a DC-DC booster converter from a 5 V USB input. The SiPM gain can reach upto 3$\times10^6$, while its typical dark count rate is 1200 kHz. The exponential decay of the generated pulse is less than 1 $\mu$s.

The SiPM is connected to the rest of the voltage processing circuit and the plastic scintillator block is wrapped in reflective foil and insulating tape. The bare detector is mounted in a light-tight aluminum casing  ($6.6\times7.4\times4.0$ cm$^3$). The main printed circuit board (PCB) amplifies and shapes the SiPM signal that is then processed by a micro-controller, an Arduino Nano. Its configuration script \cite{githubCW} allows us to control the OLED screen and data acquisition conditions and display data in real time.

The detector is powered through a mini-USB to USB connector. An event is recorded, if the signal passed a given threshold of 35 ADC (analog-to-digital converter), which is equivalent to approximately 25 mV. The average deadtime per registered event is 15 ms. A file with each event's recorded data, including its pulse amplitude, time  and deadtime, can be written using a Python script. 

\section{Simulations with Geant4}
\label{sec:simulation}

\begin{figure}[htb]
\centering
\includegraphics[scale=3]{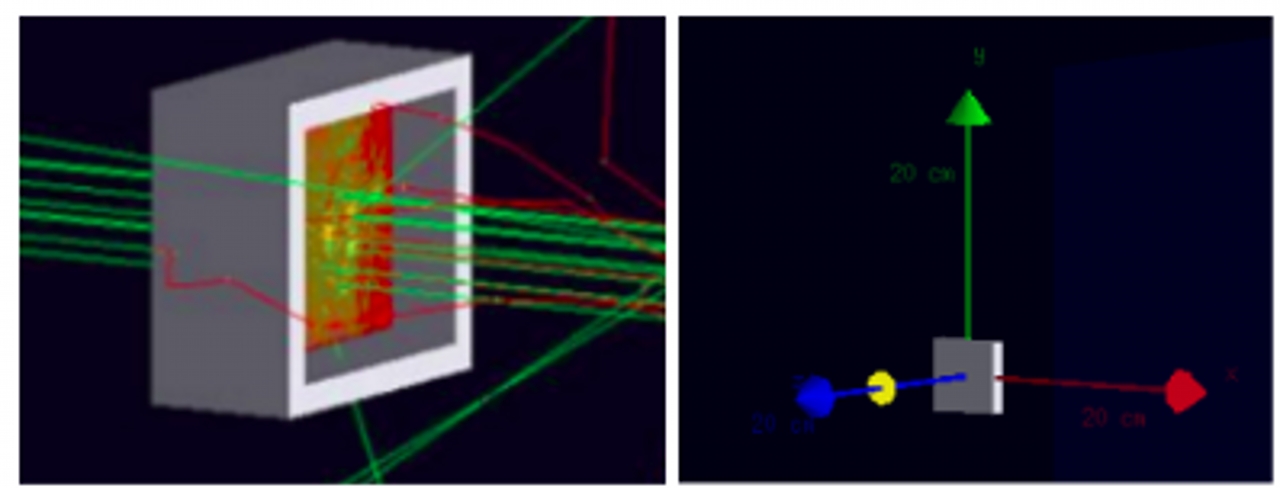}
\caption{Geant4 DMD simulation geometry. Left: perpendicular muon beam, Right: cylindrical radioactive source and DMD.}
\label{fig:sim1}
\end{figure}

In order to compare to measurements, we have developed simulations in Geant4 v10.03 \citep{Agostinelli:2002hh}, a CERN toolkit that simulates the passage of particles through matter.
The simulation scripts can be found in the supplementary material \cite{github}. We simulate the geometry of the DMD (Fig. \ref{fig:sim1}) as a plastic scintillator (i.e. BC408 plastic) with a complete description of its optical and electromagnetic properties (refraction index, absorption length, scintillation, wavelength shift properties, etc.), along with the silicon sensor  inside the aluminum casing. The circuitry was not included in the simulation.

The following physics list was used in the simulation: electromagnetic interactions (default emstandard including G4eMultiple\-Scattering,    G4Mu\-Multiple\-Scattering and G4h\-Multiple\-Scattering), hadronic interactions (G4Hadron\-Elastic\-PhysicsHP, \texttt{G4Hadron\-Physics\-FTFP\_BERT\_HP}, G4Ion\-Elastic\-Physics, G4Ion\-Physics), decay physics (G4Decay\-Physics, G4Radioactive\-Decay\-Physics), as well as optical physics (G4Optical\-Physics) and gamma nuclear interactions (G4Photo\-Nuclear\-Process).

The criteria used to trigger an event count is that a particle crossing the plastic scintillator generates optical photons that, after scattering, reach the SiPM's area and are detected. For every particle in the detector, we can collect information, including deposited energy and number of generated photons. Using the simulation, we calculate the pure geometrical efficiency of the configuration for photon collection at the SiPM. For the angular and energy distribution of atmospheric muons this efficiency is on average 85\%. In addition, we include the SiPM wavelength-dependent quantum efficiency \citep{sensL}, which is 41\% at its peak, 420 nm. Given $n$ photons that reach the SiPM, then the geometrical efficiency is modified by the combined photon detection probability: $1-(1-p)^n$, where $p$ is the individual photon wavelength-dependent quantum detection probability. The SiPM efficiency lowers the total efficiency to 76\%.

We use the theoretical atmospheric muon flux, including its angle and energy dependence \citep{Shukla:2016nio}, as the injected particle distribution. Since the flux depends on the location, we use the flux at Lima, Peru's coordinates and altitude, where the measurements were performed. For the radioactive source activity calculations, we simulate a cylindrical source of height 0.3 cm and radius 1.1 cm, as shown in Fig. \ref{fig:sim1}, at a given distance from the DMD and force its decay. We also model simple barriers and large structures (e.g. a 4-storey building) to simulate cosmic ray attenuation. 
 
\begin{figure}[h!] 
\centering
\begin{subfigure}{.6\textwidth}
 \centering
 \includegraphics[width=.99\linewidth]{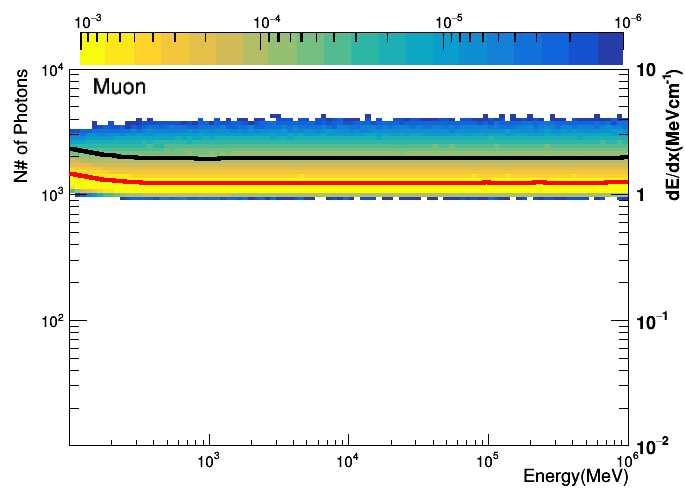}
  \end{subfigure}
\\
\begin{subfigure}{.49\textwidth}
 \centering
  \includegraphics[width=.99\linewidth]{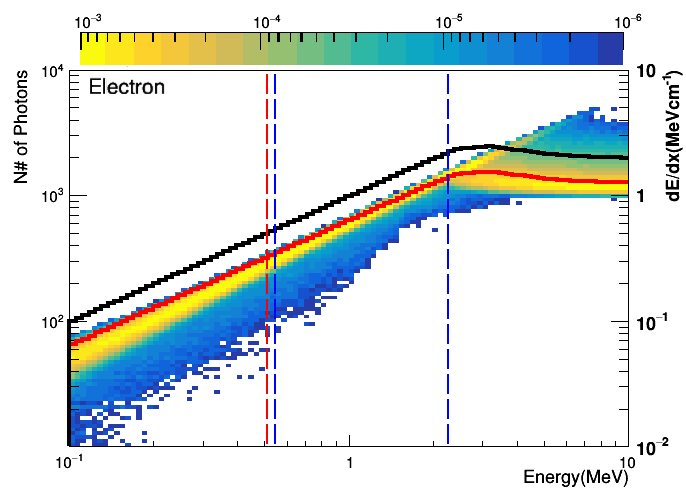}
  \end{subfigure}
    \begin{subfigure}{.49\textwidth}
  \centering
  \includegraphics[width=.99\linewidth]{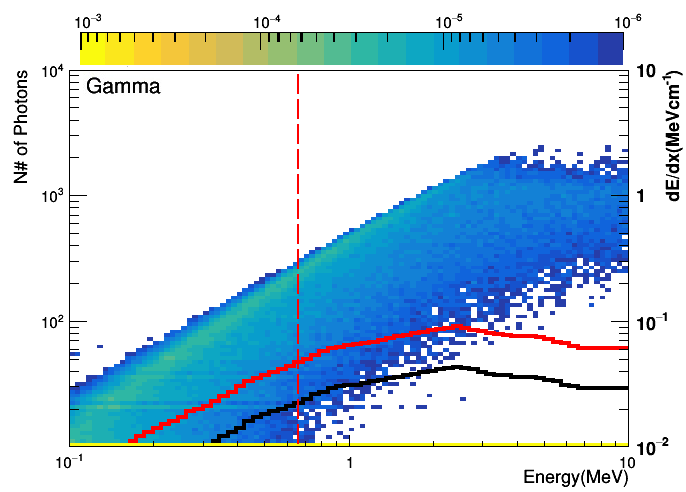}
   \end{subfigure}
\caption{Normalized 2D distribution of the simulated number of photons produced in the scintillator versus the energy of vertically incident muons (top), electrons (bottom left), and gamma rays (bottom right). The black line shows the deposited $dE/dx$ in the scintillator, the red line the photon production reconstructed from Birks' law using the $dE/dx$ and the dashed lines the maximum $\beta$ and $\gamma$ energies for $^{137}$Cs in red, and for $^{90}$Sr and $^{90}$Y in blue.}
\label{fig:emulsim}
\end{figure}
 
In Fig. \ref{fig:emulsim} we show the simulations' results for the optical photon production and $dE/dx$ (shown in black) by muons, electrons and gamma rays as a function of their energy. Photon production in a scintillator can be described by Birk's law \citep{Birk}:

\begin{equation}
\label{eq:birk}
\frac{dL}{dx}  = S \frac{\dfrac{dE}{dx}}{1+k_B\dfrac{dE}{dx}}
\end{equation}
where L is the light yield, S represents the scintillation efficiency, $k_B$ is Birk's constant and $dE/dx$ the mean energy loss per path length of the particle. For large $dE/dx$ the equation behaves nonlinearly. 

In the case of muons, the mean energy loss can be obtained with the Bethe-Bloch formula \citep{Bethe}. If this energy loss is taken from the simulation and used in Eq. \ref{eq:birk}, the resulting photon production as a function of energy, shown with a red line in Fig. \ref{fig:emulsim}, can explain the simulation results. Muons show Minimum Ionizing Particle (MIP) behavior very early in their energy spectrum, with 1300 photons produced on average. 

Electrons, on the other hand, deposit most of their energy and have a linear photon production up to $\sim$ 3 MeV, meaning that electrons of two different energies can be distinguished from their number of generated photons. This behaviour will be later useful for differentiating radioactive sources. In fact, we are highlighting in Fig. \ref{fig:emulsim} the energies of the two radioactive sources that we will later use, which are producing distinguishable amounts of photons. At energies higher than 3 MeV, electrons start showing MIP-like behavior, where their average photon production is also around 1300 photons. 

Low-energy gamma rays produce almost no optical photons in the scintillator, as seen in Fig. \ref{fig:emulsim}. The underflow bin, corresponding to zero photons, has the largest density. At the plotted energies the main photon production mechanism is Compton scattering. As shown, there is a relatively low probability of emitting hundreds of photons, thus the average number of produced optical photons (red line) is hardly increased.
  
\section{Detector Calibration with atmospheric muons}
\label{sec:calibration}

DMD’s are rather novel technology, and even though their structure is simple, their behaviour has been not fully characterized. Due to slight differences in the circuitry and SiPM to plastic coupling, a DMD may have a higher sensitivity or higher background noise, producing differences in the measured event rate. Therefore, an absolute calibration is needed.

\begin{figure}[htb]
\centering
\includegraphics[scale=0.25]{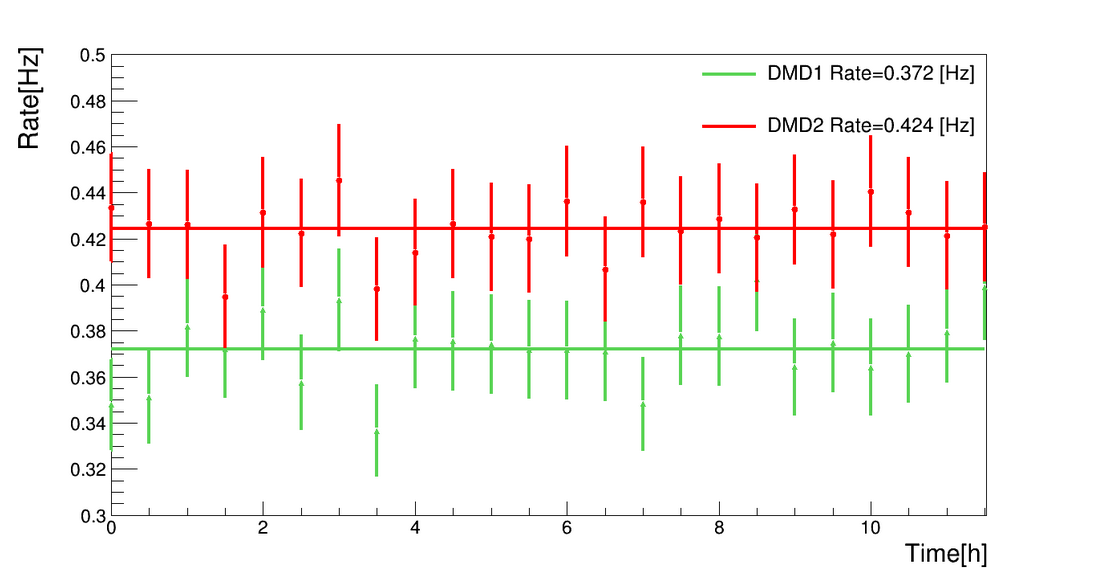}
\caption{Rate (averaged every 30 minutes) for single detectors at 60 m  above sea level at 12\degree S  77\degree W. The errors shown include statistical and systematic uncertainties. The rate time dependence is similar for both detectors.}
\label{fig:rate12}
\end{figure}

Detectors have similar time dependence for the same kind of measurement (Fig. \ref{fig:rate12}), for a short period of time, where no major atmospheric, seasonal or solar effects are relevant. Thus we prove the stability of the detectors. 

There are no corrections for temperature, since large daily temperature variations in Lima are not common (maximum of 6 $\degree$C) and the measurements were done mostly in the same period during the day. In addition, the temperature effect in the dark current, estimated \citep{sensL} for a total bias voltage of 29.4 V in the 14 $\degree$C to 26 $\degree$C range, is approximately 3.5 $\mu$A. The equivalent 3.5 mV change in our circuit is below the set signal threshold of 25 mV.

The total expected muon flux is 1.16 cm\textsuperscript{-2}min\textsuperscript{-1}, based on an average rate calculated with PARMA/EXPACS \citep{expacs}, using a 60 m altitude and different parameters according to the registered solar activity at various stages of the solar cycle. The expected total rate for a DMD is 0.484 s\textsuperscript{-1}, resulting in a measured 76.8\% and 87.6\% efficiency for the two used detectors. As described in Sec. \ref{sec:simulation}, the simulated geometrical efficiency including the quantum efficiency of the detector for atmospheric muons is 76\%. Additional events could be attributed to the circuit's background noise or a not complete light-tight enclosure.

\begin{figure}[htb]
\centering
  \includegraphics[scale=1.5]{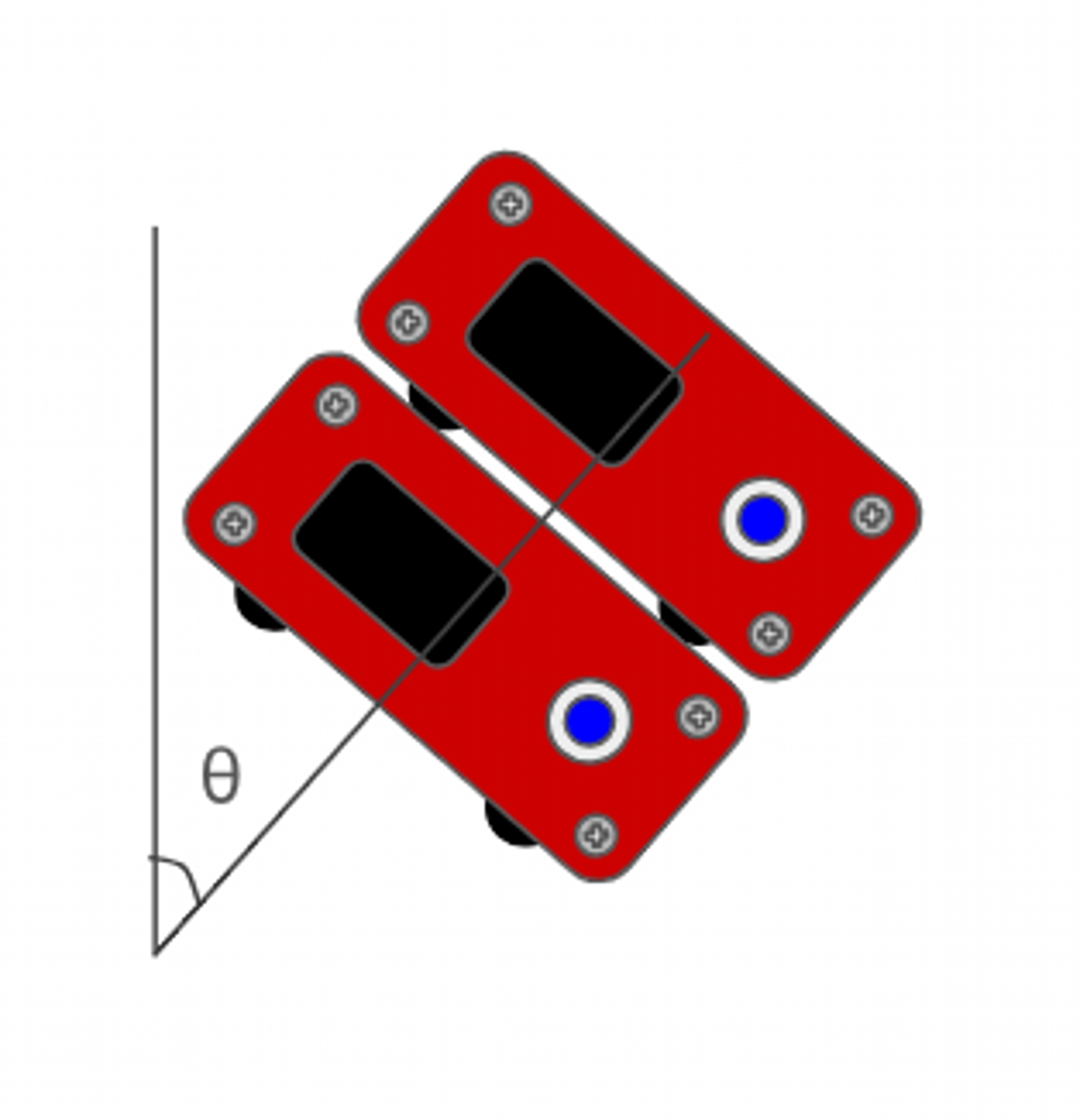}
\caption{Setup for coincidence measurements of the atmospheric muon angular distribution. The corresponding angular aperture of this configuration is 8.9$\degree$.}
\label{fig:setupang}
\end{figure}

\subsection{Angular distribution}

The first check of the detector's performance is done with the  angular distribution of the atmospheric muon flux. We performed coincidence measurements, using two DMDs placed one on top of the other, as shown in Fig. \ref{fig:setupang}, to reduce the aperture uncertainty. 

We use an offline trigger with time gap of less than 0.01 s to count  measurements in both DMDs as a single coincident event. This time gap gives an accidental coincidence rate of 7.22 mHz, which represents less than 8\% of the measured coincidence rate. The accidental coincidence rate is subtracted in the results. Measurements were taken for ten hours outdoors on the rooftop of the Physics Section building, away from all possible shielding to avoid attenuation effects. The measured angles were at 0\degree, 22.5\degree, 45\degree, 67.5\degree, and  90\degree with respect to the zenith. At 0\degree ~the coincidence rate is 0.095 Hz.
	
\begin{figure}[htb]
\centering

\begin{subfigure}{.49\textwidth}
 \centering
  \includegraphics[width=.99\linewidth]{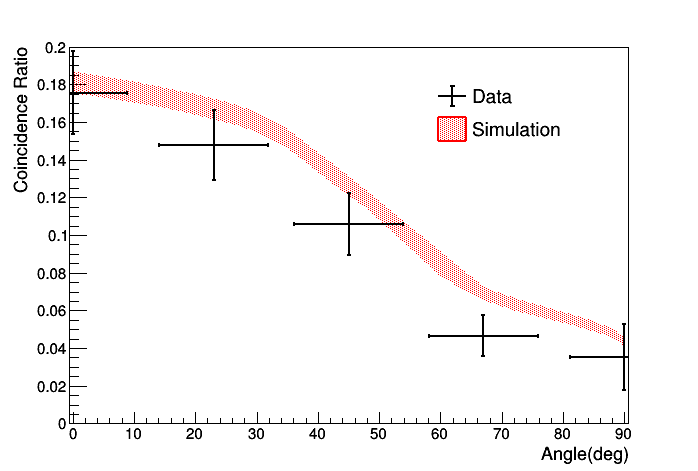}
  \end{subfigure}
    \begin{subfigure}{.49\textwidth}
  \centering
  \includegraphics[width=.99\linewidth]{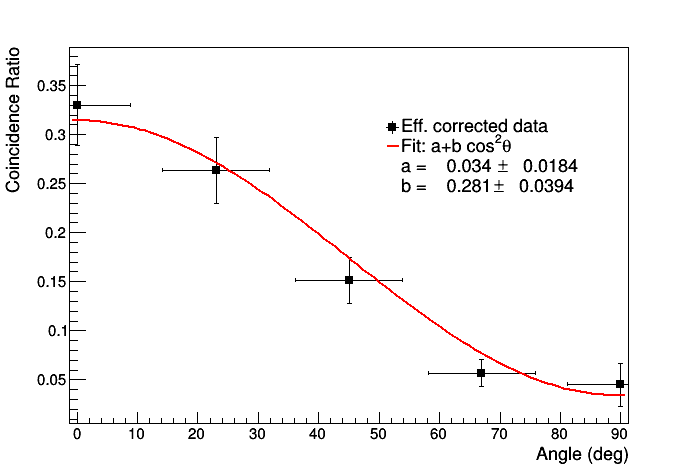}
   \end{subfigure}

 \caption{Left. Angular distribution of measured atmospheric muons coincidence event ratio obtained using two DMDs (black crosses). The ratio is obtained dividing the number of coincident events by the events measured in only one detector. The measurements were done outdoors on the rooftop of the Physics Section building, away from any shielding. The red band, including errors, is the simulation  corrected for the measured efficiency at different angles. The simulation has as input the theoretical muon distribution which follows $\cos^2(\theta)$. The error bar on the angle axis represents the geometrical aperture of the detectors configuration, while the vertical  error stands for the statistical and systematic errors added in quadrature. The comparison between data and simulation gives a $\chi^{2}=4.13$ with 5 degrees of freedom and fit probability of 53\%. Right. Angular distribution of the efficiency corrected measured atmospheric muons coincidence event ratio obtained using two DMDs (black crosses). The red line is the fit to the data using the theoretical prediction $A+B\cos^2(\theta)$, with $\chi^{2}=0.97$ with 3 degrees of freedom and fit probability of 81\%.
 }
 \label{fig:atmmuon}
\end{figure}

In Fig. \ref{fig:atmmuon} we show the comparison between measurements and simulation. In this case, the simulation has two additional corrections: one for the relative efficiency of each detector and one for the actual detector efficiency for each angle. In order to have an independent measurement of the efficiency at each angle a configuration of three detectors one on top of the other is used. We expect that if the two outer detectors have a coincidence, the inner detector must also present a signal. Then the experimental efficiency is the fraction of events with a signal in the inner detector when the outer detectors are trigger in coincidence divided by the times the two outer detector have a coincidence. This efficiency varies from 75$\pm4$\% at 0$\degree$ to 95$\pm23$\% at 90$\degree$. The difference between simulated and measured efficiency, as well as the difference between detector efficiencies are added as systematic errors to the data points. The measured data are in agreement with the simulation with a 53\% probability. Muon's energy loss and decay depend on the height of their production layer and amount of material traversed \citep{Forbush}. The muon flux, then, decreases as the zenith angle increases, since at larger angles there is a higher chance of interaction in the atmosphere due to the longer distances muons must travel to reach the detector. The measured distribution including efficiency corrections fits with probability 81\% the theoretical distribution $A+B\cos^2(\theta)$, as seen in the Fig. \ref{fig:atmmuon}.

\begin{figure}[htb]
\centering
  \includegraphics[scale=3]{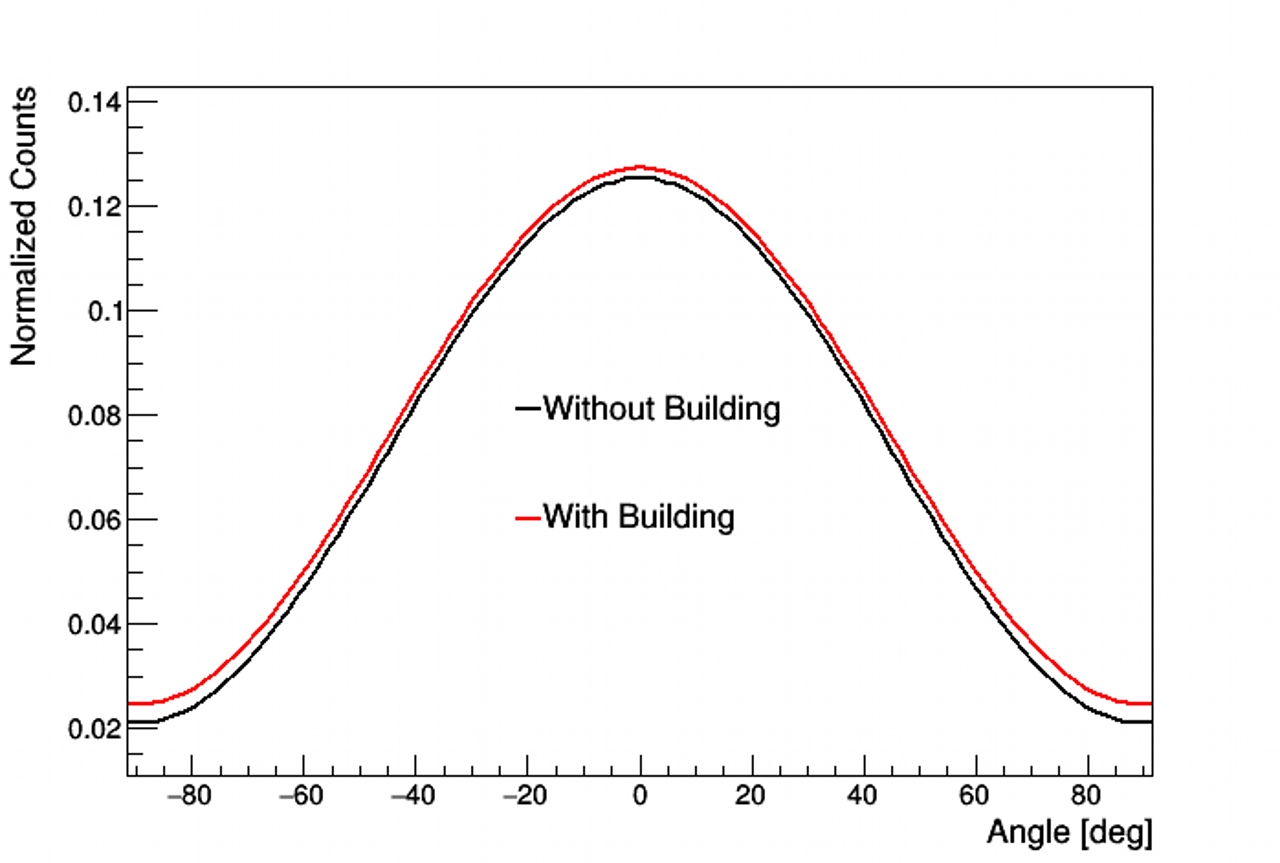}
\caption{Simulation of atmospheric muon angular distributions comparing building attenuation, normalized to the total number of generated events in Geant4.}
\label{fig:bmuon}
\end{figure}

\subsection{Attenuation}
Next we look for attenuation effects due to large structures on the atmospheric muon flux. We simulate the Physics Section building, where we made our measurements. The toy simulation is a four story building, with an area of 4 m$^2$, built of concrete. We considered an average density of 1.5 g/cm\textsuperscript{3} and do not include windows, beams, nor columns.

As seen in Fig. \ref{fig:bmuon}, the simulated angular distributions without and including the building are alike,  suggesting that the atmospheric muons are much not deflected and do not interact with the building. 

As a further analysis, we study the effect of muon attenuation in four settings: (1) the detector on an open field, (2) on the open field with a single 1.5 cm lead plate, (3) alone inside the previously described building, and (4) with the lead barrier inside the building. First, we considered a comparison between the detector on an open field and shielded with the lead plate. For the simulation, muons are directed towards a 10$\times$10 m\textsuperscript{2} area (much larger than the detector) and a million events are generated. The measurements, in all four settings, were two hours each. The ratio, Table \ref{table:1}, is almost one in both cases (simulation and measurement). There is no difference between the measured rate at an open field and below an 1.5 cm thick lead plate. 

We then considered the detector without the lead plate inside the four storey building and on an open field. The results, Table \ref{table:1}, were once again very close to one. We can then conclude that separately, both the lead plate and the building, produce no visible attenuation.

\begin{table}[htb]
\centering
\begin{tabular}{ |l | c | c | c | } 
 \hline
  &  $\frac{\mbox{lead}}{\mbox{open field}}$ & $\frac{\mbox{building}}{\mbox{open field}}$ & $\frac{\mbox{building+lead}}{\mbox{building}}$\\
 \hline
 Measurement & ${ 0.99 \pm 0.04}$ & ${0.99 \pm 0.04}$ & ${0.87 \pm 0.04}$\\ 
  \hline
 Simulation & ${1.01 \pm 0.01}$ & ${1.01 \pm 0.06}$ & ${0.93 \pm 0.06}$\\
 \hline
\end{tabular}
\caption{Ratio of events with the detector in different configurations. Measurements lasted for two hours each and a million events were generated in the simulation. The measurement errors included statistical and systematic uncertainties.}
\label{table:1}
\end{table}

Finally, we considered a combined measurement (lead plate and building). The ratio of the combined configuration to the building alone one, Table \ref{table:1}, gives a value that is slightly lower than one but still compatible with unity within 3.3 $\sigma$. This results is in agreement with simulation within 0.8 $\sigma$. Looking in the MC truth at the particles that reach the detector (Table \ref{table:2}), we  see that the building, compared to the open field, is creating more (4 $\sigma$) secondary particles (electrons) per event that produce photons in the scintillator. However, the number of secondary electrons that reach the plastic scintillator diminishes to some extent, with a 1.6 $\sigma$ difference, when the lead plate is added. 

\begin{table}[htb]
\centering
\begin{tabular}{|l|c|c|c|c|} 
 \hline
  &  Open field & Building & Lead & Combined \\
    \hline
 Events & ${ 525 \pm 23}$ & ${528 \pm 23}$ & ${531 \pm 23}$ & ${490 \pm 22}$\\ 
  \hline
 No $\mu^{-}$ & ${52 \pm 7}$ & ${65 \pm 8}$ & ${67 \pm 8}$ & ${74 \pm 8}$ \\
 \hline
$e^-$/event & ${0.27 \pm 0.03}$ & ${0.47 \pm 0.04}$ & ${0.34 \pm 0.03}$ & ${0.39 \pm 0.03}$ \\
 \hline
\end{tabular}
\caption{Simulation of charged particles in the detector. Combined refers to simultaneous use of the building and lead plate. Events without muons (No $\mu^-$) are events with only electrons. The simulation is based  on one million generated events.}
\label{table:2}
\end{table}

\section{Source Activity and Identification}
\label{sec:radioactive}

Even if the original design of the DMD was made for muon identification, we are going to prove next that it can also be used to measure charged particles, such as electrons. For this purpose we make measurements of radioactive beta sources using the DMD. 

It is important to note that for measuring electrons it is required to remove the DMD's aluminum shielding. For isolating the radioactive source signal we need to subtract the atmospheric muon background. Then, we perform simultaneous measurements with two DMDs. From now on, we will call them DMD1 and DMD2. DMD1 is directly exposed to the $\beta$ source, giving $R_{S+B}$, the signal plus background rate. DMD2 is kept as control, measuring the muon background rate, $R_{B}$. The simultaneity of the measurements is due to possible time variations in the muon atmospheric background, producing two different backgrounds if we made the measurements at different times with a single detector. In order to make comparable the measurements of the muon background rates of these two different detectors, we make simultaneous measurements of the aforementioned rates obtaining a relative efficiency factor that we will call $\eta$, defined in Eq. \ref{eq:eta}. This factor allows us to convert the rates between the two detectors. 
We can then obtain the pure signal rate, R$_{S}$ (see Eq. \ref{eq:puresignal}), by subtracting the atmospheric muon background counted with the control DMD2. 

\begin{equation}
\label{eq:eta}
\eta =\dfrac{R_{B}^{\mbox{DMD1}}}{R_{B}^{\mbox{DMD2}}}
\end{equation}

\begin{equation}
\label{eq:puresignal}
R_{S}^{\mbox{DMD1}}=R_{S+B}^{\mbox{DMD1}}-\eta\times R_{B}^{\mbox{DMD2}}
\end{equation}

\begin{figure}[htb]
\centering
\includegraphics[scale=3, trim=0 15 0 0,clip]{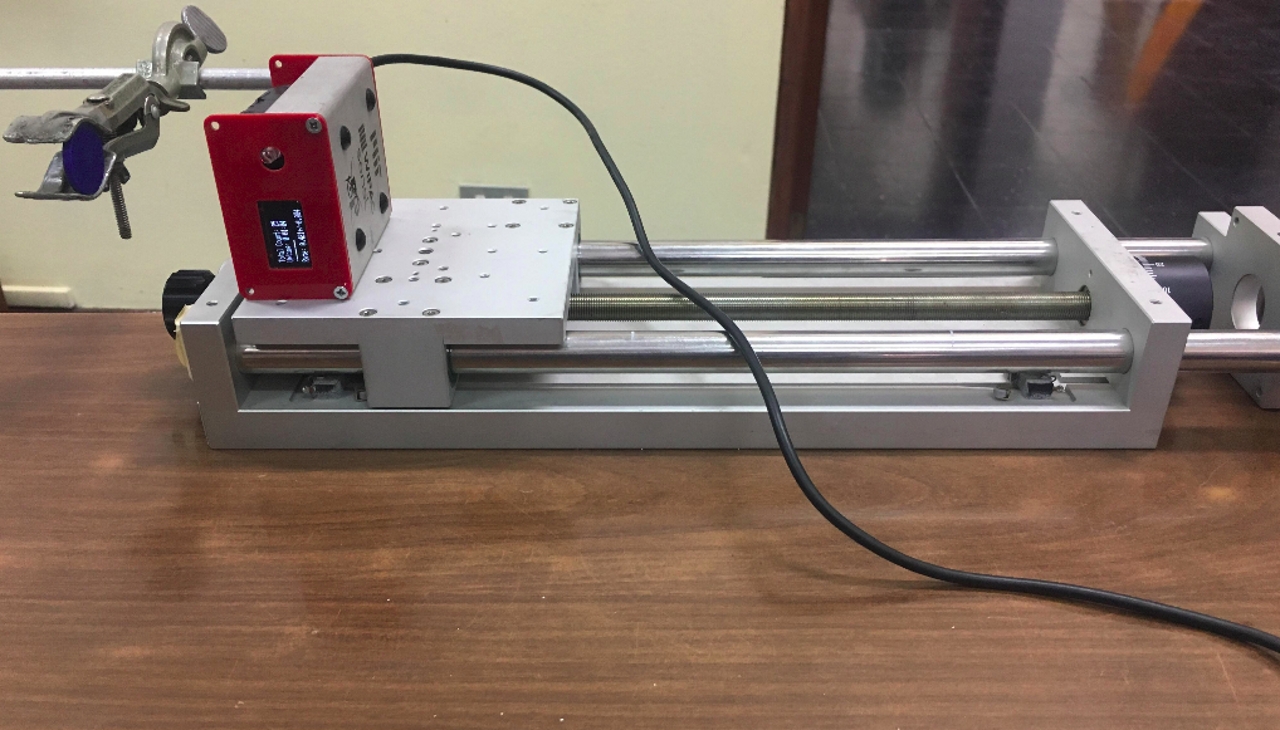}
\caption{Setup for radioactive source measurements at fixed distances. The DMD has the half of its aluminum case facing the source removed. The DMD is placed on top of the precision platform, while the radioactive source disk is hold fixed.}
\label{fig:setup2}
\end{figure}

\begin{figure}[htb]
\centering
\includegraphics[scale=0.45]{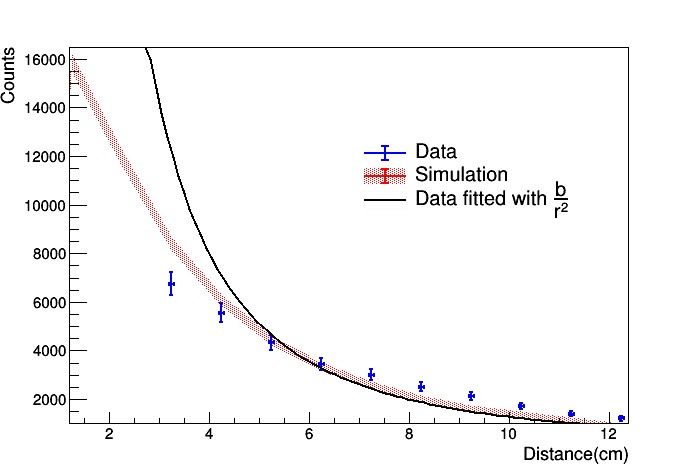}
\caption{DMD measured and simulated counts as a function of distance ($r$) from a $^{90}$Sr source.  Comparing data with simulation gives a $\chi^{2} = 16.9$ with 9 degrees of freedom and fit probability 10\%. Data points are also fitted with an inverse squared law ($b\frac{1}{r^2}$), with a worse $\chi^{2}/NDF = 170/9$.}
\label{fig:linear}
\end{figure}

To maintain a standard distance between source and DMD we used a precision platform. This setup is shown in Fig. \ref{fig:setup2}. We did an experimental cross-check to see that the source intensity diminishes with the square of the distance as seen in Fig. \ref{fig:linear}. The inverse squared distance proportionality from the point source approximation starts to deviate from measurements and simulations for distances smaller than 5 cm because of the source dimensions (a cylindrical slab of 2.2 cm diameter and 0.3 cm height, with actual radioactive diameter of 0.63 cm) and collection area (5 cm $\times$ 5 cm plastic scintillator). 

\subsection{Activity measurement}

We used two radioactive isotopes for this study: $^{90}$Sr and $^{137}$Cs. Both decay through pure $\beta$ emission. $^{90}$Sr ($t_{1/2}=28.79$ yr) emits an electron ($E_{max}=0.546$ MeV) going to $^{90}$Y, which is also a $\beta$ emitter ($E_{max}=2.28$ MeV, $t_{1/2}=64$ hours). $^{137}$Cs ($t_{1/2}=30.17$ yr) emits an electron ($E_{max}=0.514$ MeV) going to $^{137m}$Ba, which is a $\gamma$ emitter ($E=0.66$ MeV, $t_{1/2}=153$ s). As a result, $^{137}$Cs has effectively a large $\gamma$ component. Both beta sources emits betas with very similar energies. However, Y-90, the daughter of Sr-90, has a very different energy. Thus, we will be able to distinguish the $^{90}$Sr source by its daughter. Since their lifetimes are quite different, it will be in practice the same measurement. 

We measured the background subtracted rates (see Eq. \ref{eq:puresignal}) of two sources of $^{90}$Sr and two sources of $^{137}$Cs, each with a different activity as shown in Table \ref{table:4}. From this point onward, rate refers to the background subtracted rate. The lower activity sources are referred to as A, while the higher ones as B. We made two hour measurements at 5 cm for $^{90}$Sr sources and at 0.5 cm for $^{137}$Cs. The closer distance for $^{137}$Cs is due to its lower rate, which at the larger distance was compatible with the muon background. 

The activities at the time of observation are calculated from their initial activities and their production year. $^{137}$Cs rates are lower than $^{90}$Sr, in spite of their respective activities being higher. This is because $^{90}$Sr, through the $^{90}$Y decay, produces more energetic electrons, which emit more scintillation photons (see Fig. \ref{fig:emulsim}). On the other hand, gammas from $^{137m}$Ba, product of $^{137}$Cs, are practically invisible to the DMDs, as simulated in Fig. \ref{fig:emulsim}. This has been corroborated by testing the DMD's response to a $^{55}$Fe source ($t_{1/2}=2.74$ yr, activity 1 $\mu$Ci) which is an X-ray (5-6 keV) and Auger electron (5.19 keV) emitter. At these energies there is neither photon, nor electron detection.

The activities can be extrapolated from the DMDs measured rates (Table \ref{table:4}) and then compared with the \textit{actual} ones. In order to estimate the activity we obtain first a conversion factor ($\alpha$=Activity$_A$/Rate$_A$), which depends on the isotope, the detector and the distance. Then the activity for other sources of the same isotope can be calculated using this conversion factor and the measured rate with the same detector and distance (Activity$_B$=$\alpha$ Rate$_B$). The extrapolated activities of source B obtained with the conversion factor from source A have less than a 15\% error and are in both cases ($^{90}$Sr and $^{137}$Cs) within 0.6 $\sigma$ of the known activity, as shown in Table \ref{table:4}. 

\begin{table}[htb]
\centering
\begin{tabular}{| c | c | c | c |} 
 \hline
 Source & Rate(Hz) & \multicolumn{2}{c|}{Activity($\mu$Ci)} \\ 
        & & expected     & extrapolated$\pm$stat$\pm$sys \\ 
 \hline
 $^{137}$Cs A & ${ 1.01 \pm 0.1}$& ${ 0.23 \pm 0.02}$  &\\
  \hline
 $^{137}$Cs B  & ${ 6.53 \pm 0.65}$ & ${1.77 \pm 0.40}$\footnote{This activity was obtained using a GAMMA-SCOUT\texttrademark since the source's production year was unknown. The error on the measurement is included.}  & ${1.49 \pm 0.07 \pm 0.22}$\\
 \hline
 $^{90}$Sr A & ${ 6.16 \pm 0.04}$ & ${0.065 \pm 0.004}$ & \\
 \hline
 $^{90}$Sr B & ${ 9.16 \pm 0.06}$ & ${0.093 \pm 0.005}$  & ${0.097 \pm 0.003 \pm 0.005}$\\
 \hline
\end{tabular}
\caption{Beta sources background subtracted rates measured for two hours, expected and extrapolated activities. The $^{90}$Sr ($^{137}$Cs) measurements were done at a distance of 5 cm (0.5 cm) from the detector. The expected activities are calculated from their initial factory activities (calibrated with a 5\% error) and production year. A further uncertainty is added taking a one year window, since only the purchase year is known, which represents a 3.2\% error on average. The extrapolated activities show separated statistical and systematic errors, while for the other errors they are added in quadrature.}
\label{table:4}
\end{table}

\subsection{Beta source identification}

\begin{figure}[htb]
\centering
\includegraphics[scale=3]{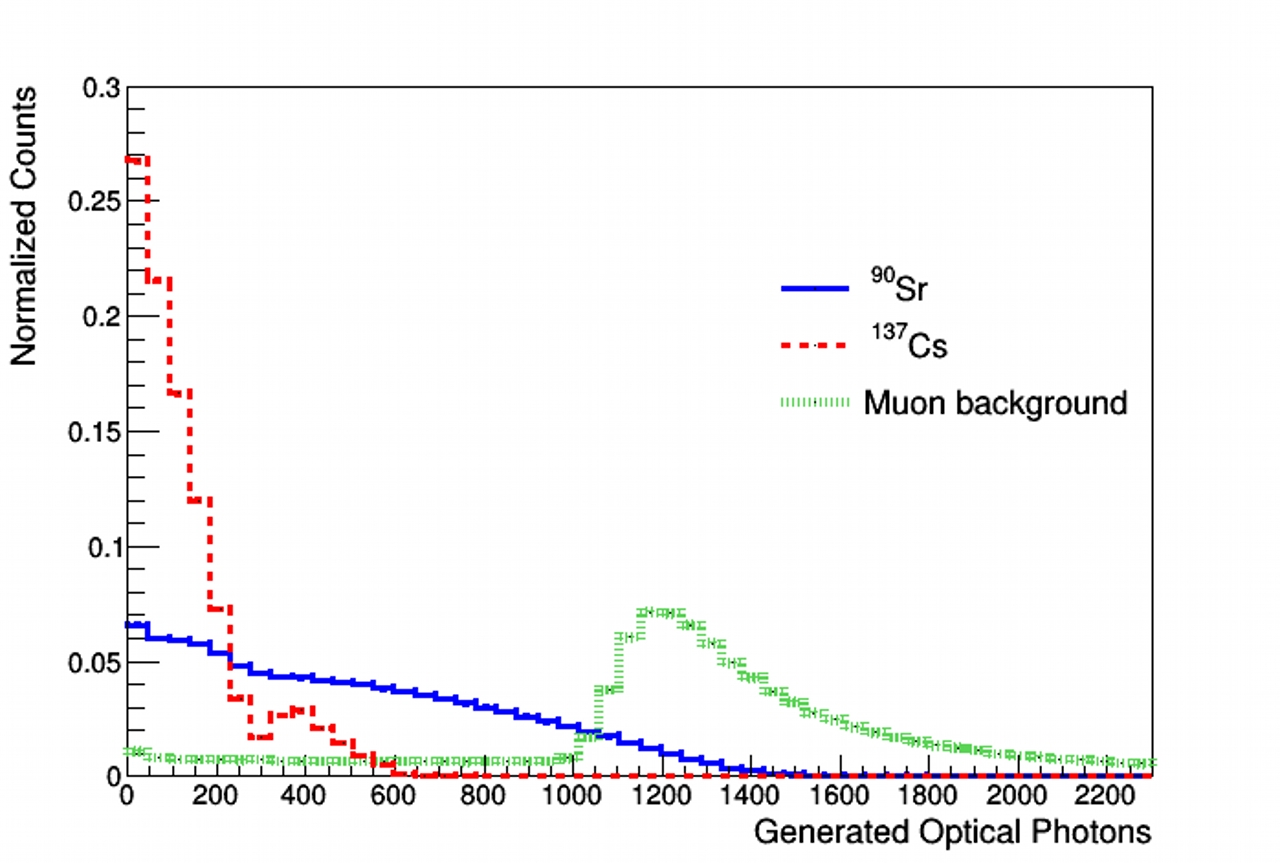}
\caption{Normalized optical photon production comparison for simulated beta sources and muon background.}
\label{fig:comp4}
\end{figure}

We have developed Geant4 simulations showing that $\beta$-sources identification is possible using a plastic scintillator. A difference in the produced optical photon spectra between atmospheric muons and radioactive beta sources must reflect on their respective measured voltage spectra, as seen in Fig. \ref{fig:comp4}. In this way, we can identify radioactive $\beta$-sources. 

\begin{figure}[htb]
\centering
\includegraphics[scale=0.45]{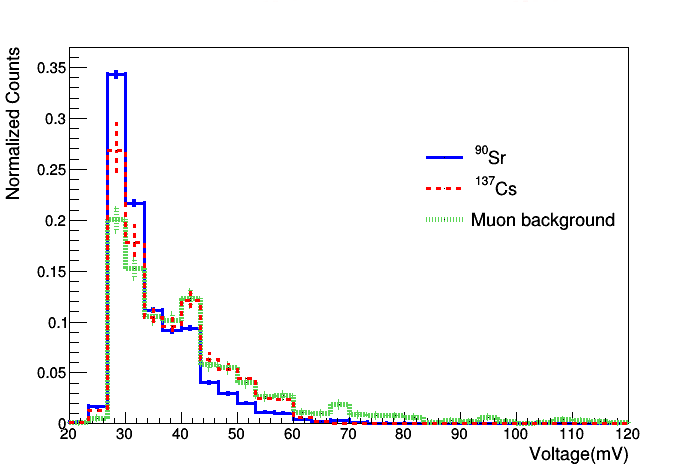}
\caption{Normalized measured SiPM voltage spectra for the atmospheric muon background and background-subtracted radioactive sources with the highest activity. Events with voltages $>70$ correspond almost exclusively to the muon background. Error bars include statistical and systematic errors.}
\label{fig:voltage}
\end{figure}

We then analyze the voltage spectrum for each source, shown in Fig. \ref{fig:voltage}. Because of slight differences in the assembly process, the voltage spectrum is different for each detector, so the same DMD must be used for every measurement. In this case, in order to subtract the atmospheric muon background, preserving the corresponding detector voltage spectrum, we use the following equation:

\begin{equation}
\label{eq:spectrum}
R_{S}^{\mbox{DMD1}}(V_i,t)=R_{S+B}^{\mbox{DMD1}}(V_i,t)-\left(\frac{R_{B}^{\mbox{DMD2}}(t)}{R_{B}^{\mbox{DMD2}}(t_{0})}\right) R_{B}^{\mbox{DMD1}}(V_i,t_{0})
\end{equation}

where $V_i$ is the voltage bin, $t$ and $t_0$ represent the times at which the simultaneous measurements where performed. Then, for a given voltage bin the resulting signal rate at time $t$ with DMD1 will be the measured rate of DMD1 at $t$ minus the normalized background rate from DMD1 measured at a previous time $t_0$. Thus, we keep the spectrum of the same detector. The normalization is obtained by comparing measurements of only background with DMD2 at $t$ and $t_0$. These last measurements are voltage integrated, thus do not depend on the spectrum.

A 25 mV threshold is applied in order to remove the noise from the DMD. Because of this, the peak from $^{137}$Cs corresponding to the least energetic events seen in Fig. \ref{fig:comp4} is discarded.

We used two-sample Kolmogorov$-$Smirnov (KS) tests to assess the level of similarity of the voltage distributions to prove if their spectra were distinguishable from the atmospheric muon background and from those of different radioactive isotopes. 

We applied the following procedure, based on a Monte Carlo approach, to calculate the KS results including errors. We turn each SiPM voltage histogram bin into a PDF using either a Poisson or Gaussian distribution depending on the bin value. Hence, we have as many PDFs as bins in the histogram for each dataset. Then we randomly generate 10000 normalized distributions for each data set according to their individual PDFs. We calculate their CDFs and apply the KS test for each pair of the newly generated distributions, to find the D parameter (i.e. maximum distance between distributions). The final D parameter is the average of each individual D from the MC process with the standard deviation as its error. 

\begin{figure}[!htb]
\centering
\includegraphics[scale=0.45]{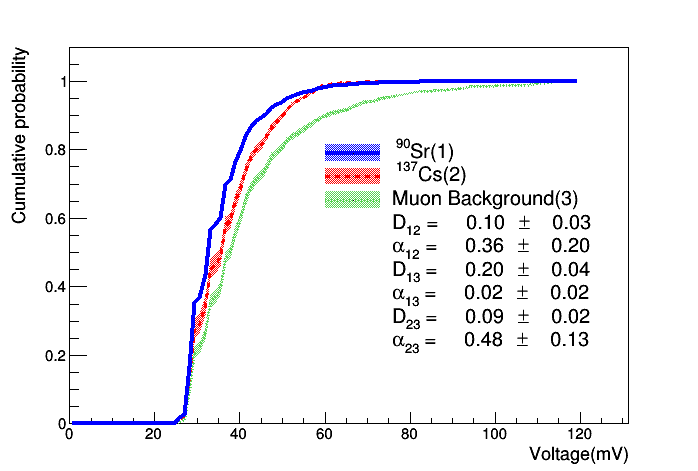}
\caption{Kolmogorov$-$Smirnov tests showing the cumulative probability comparing radioactive sources and muon background. The D parameter represents the maximum distance between distributions and $\alpha$ is the significance level. The sample size (i.e. number or bins) is 100.The error bands were obtained by taking the  standard deviation from the 10000 generated CDFs.}
\label{fig:ks}
\end{figure}

To compare two sources the significance level $\alpha$, presented in Fig. \ref{fig:ks}, is then calculated. For instance, if $\alpha<0.003$, we can reject at the 3$\sigma$ level the null hypothesis that the two distributions are drawn from the same distribution (e.g. the $^{90}$Sr spectrum is different from the muon background). For $^{90}$Sr we conclude that the source is different from background at $\approx2.4\sigma$ level, while at $\approx0.7\sigma$ it is different from $^{137}$Cs. In the case of $^{137}$Cs, its spectrum is at $\approx0.9\sigma$ different than that from the atmospheric muon background. 

The sources of the same isotope were found to be indistinguishable, to a confidence level better than 95\%, between different samples of the same isotope ($^{90}$Sr: $\alpha$=$0.998 \pm 0.002 $, $^{137}$Cs: $\alpha$=$0.96 \pm 0.02$). 

\section{Conclusions}
\label{sec:conclusions}

We have used a new design of a portable muon detector based on a plastic scintillator with a SiPM light detector. We have calibrated and evaluated the performance of the detector by measuring and simulating the local atmospheric muon angular flux. As expected, its angular distribution follows a $\cos^2(\theta)$ function. We have also measured the attenuation to the muon flux inside a building and showed that secondary electrons are produced due to the passing of muons.

We have extended the use of the muon detector to measurements of radioactive $\beta$-sources. Using simultaneously two detectors we can obtain the background-subtracted signal rate from the sources. For the proof of concept we used two $\beta$-emitting sources: $^{90}$Sr and $^{137}$Cs, each with two disks of different activities. Our analysis was divided in two parts: finding a factor to convert measured rates into activities for each source and contrasting, through Kolmogorov$-$Smirnov tests, the voltage spectra of the sources to tell if they are similar or different. 

We found that these detectors can in fact be used to determine the source's current activity, within an error of 0.6 $\sigma$, when the signal is above the atmospheric muon background. We have also studied the feasibility for developing beta-distinguishing capabilities using this technology. In the case of $^{90}$Sr, it can be discriminated from background at the $\approx2.4\sigma$ level. However, further work is required to reach the same potential to distinguish $^{90}$Sr from $^{137}$Cs or $^{137}$Cs from background, currently at less than 0.9$\sigma$ level. For example, pulse signal characteristics can be analyzed, threshold limits tested and energy resolution assessed.

Nevertheless, gamma and alpha sources cannot be measured. Gamma sources would generate too few scintillation photons to produce a signal and alpha sources will not be able to pass the light isolation tape to arrive to the plastic scintillator. In an extension of this work other beta sources can be experimentally tested. Using simulation we find that also other beta sources as $^{151}$Sm and $^{241}$Pu have distinguishable spectra from background and other beta sources. 

\section{Acknowledgements}
This work was funded by the Direcci\'on de Gesti\'on de la Investigaci\'on at PUCP under grant No. DGI-2017-3-0019 (CAP) and DGI-2018-5-0010 (PAIN). The authors would like to thank J. Conrad, S. Axani and C. Arg\"uelles, for useful discussions, suggestions and reading of the manuscript, as well as for the detector components donations that made this work possible. We would also wish to thank the Nuclear Tracks Group for letting us use their radioactive sources and equipment, R. S\'anchez and J. A. Guerra for the measurements of the plastic scintillator's optical properties. We are also thankful to the reviewers for their comments and suggestions that helped us to improve this work.

\end{document}